\documentclass[11pt]{article}

\usepackage{graphicx}
\usepackage[square,sort,comma,numbers]{natbib} % -- needed for arxiv
\usepackage{times}
\newcommand{\bi}{\begin{itemize}\item}
\newcommand{\ei}{\end{itemize}}

\begin{document}
\title{Design and analysis of experiments linking on-line drilling methods to improvements in knowledge}
\author{
Gunnar Stefansson\thanks{University of Iceland Science Institute, Dunhaga 5, 107 Reykjavik, Iceland}
Anna Helga Jonsdottir\thanks{University of Iceland Science Institute, Dunhaga 5, 107 Reykjavik, Iceland}
}

\maketitle

\begin{abstract}
An on-line drilling system, the tutor-web, has been developed and used
for teaching mathematics and statistics. The system was used in a
basic course in calculus including 182 students. The students were
requested to answer quiz questions in the tutor-web and therefore
monitored continuously during the semester. Data available are grades
on a status exam conducted in the beginning of the course, a final
grade and data gathered in the tutor-web system. A classification of
the students is proposed using the data gathered in the system; a
\textbf{G}ood student should be able to solve a problem quickly
and get it right, the ``diligent'' hard-working \textbf{L}earner may
take longer to get the right answer, a guessing (\textbf{P}oor)
student will not take long to get the wrong answer and the remaining
(\textbf{U}nclassified) apparent non-learning students take long to
get the wrong answer, resulting in a simple classification \textbf{GLUP}.
The (\textbf{P}oor) students were found to show the least improvement,
defined as the change in grade from the status to the final exams,
while the \textbf{L}earners were found to improve the most. The
results are used to 
demonstrate how further experiments are needed and can be designed 
as well as to indicate how a system needs to be further developed
to accommodate such experiments.

\end{abstract}

\section{Introduction}
\label{introduction}

With the increasing number of web-based educational
systems several
types of educational
systems have emerged. These include learning management system (LMS),
learning content management system (LCMS), virtual learning
environment (VLE), course management system (CMS) and Adaptive and
intelligent Web-based educational systems (AIWBES).\footnote{
The terms VLE and
CMS are often used interchangeably, CMS being more common in the
United States and VLE in Europe.}

The LMS is designed for planning, delivering and managing
learning events, usually adding little value
to the learning process nor  supporting internal content
processes \cite{ismail2001design}. A VLE  provides similar
service, adding interaction with users and access to a
wider range of resources \cite{piccoli2001web}. The primary role of a
LCMS is to provide a collaborative authoring environment for creating
and maintaining learning content \cite{ismail2001design}.

Many systems are merely a network of static hypertext pages
\cite{brusilovsky1999adaptive} but adaptive and intelligent Web-based
educational systems (AIWBES) use a model of the goals, preferences and knowledge of each student and
use this to adapt to the needs of that student
\cite{brusilovsky2003adaptive}. These systems tend to be subject-specific because of their structural complexity and therefore do not provide a broad range of content.

The tutor-web (at \textit{http://tutor-web.net}) used here
is an open and freely accessible \mbox{AIWBES} system, available
to students and instructors at no cost. The system has been a research
project since 1999 and is completely based on open source computer
code with material under the Creative Commons Attribution-ShareAlike
License. The material and programs have been mainly developed in
Iceland but also used in low-income areas (e.g. Kenya).  Software is
written in the Plone\footnote{\textit{http://plone.org}}, CMS (content
management system), on top of a
Zope\footnote{\textit{http://zodb.org}} Application Server.

In terms of internal structure, the material is modular, consisting
of departments (e.g. math/stats), each of which contains courses
(e.g. introductory calculus/regression). A course can be split into tutorials
(e.g. differentiation/integration), which again consist of lectures
(e.g. basics of differentiation/chain rule).  Slides reside within
lectures and may include attached material (examples, more detail,
complete handouts etc). Also within the lectures are drills, which
consist of quiz items.  The drills/quizzes are designed for learning,
not just simple testing.  The system has been used for introductory
statistics, mathematical statistics, earth sciences, fishery science,
linear algebra and calculus in Iceland and Kenya, with some 2000 users
to date.

\begin{figure}[!h]
\includegraphics[width=14cm]{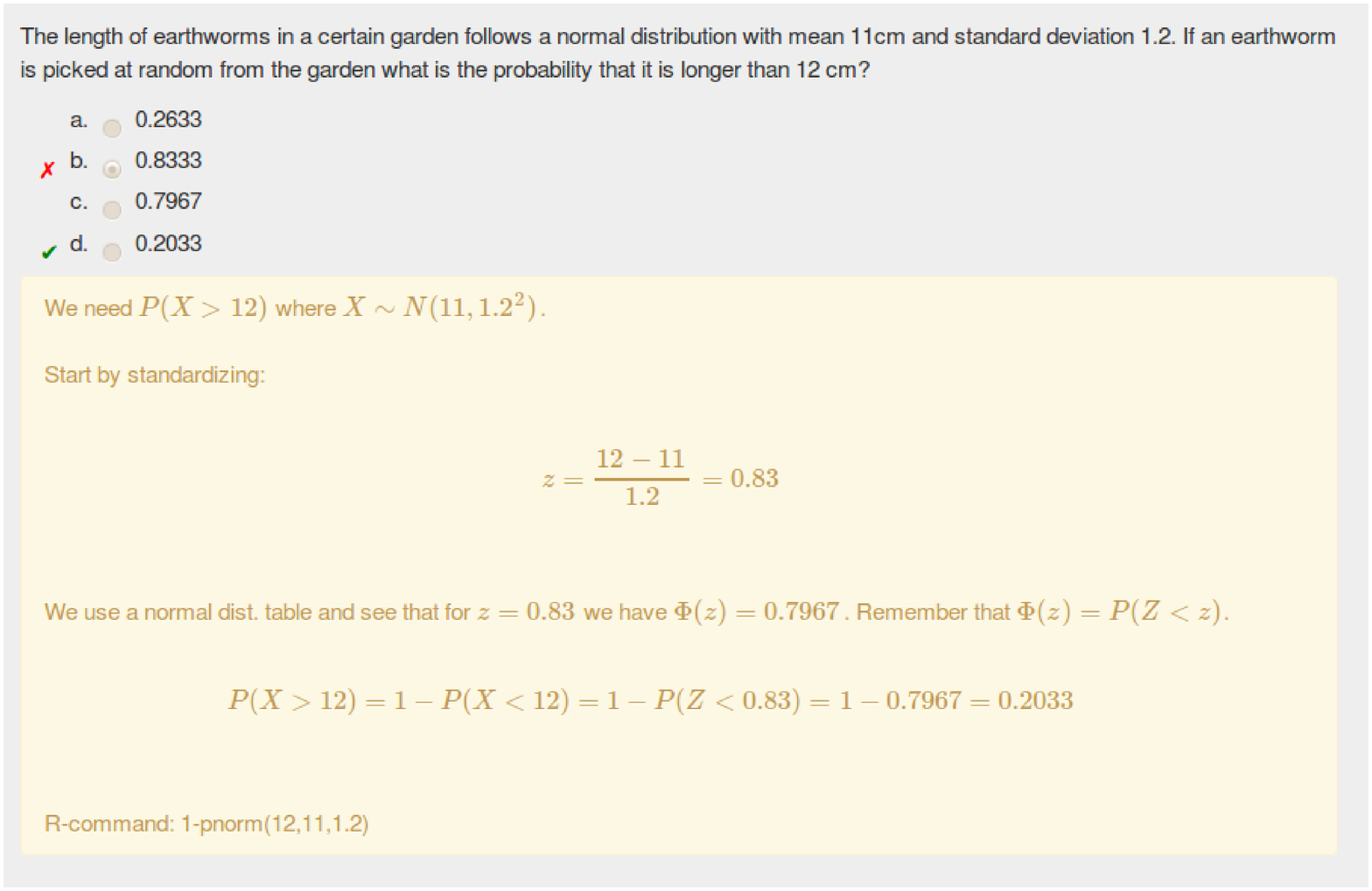}
\caption{Typical drill item, after the student has responded (incorrectly).}
\label{fig:quiz}
\end{figure}

A fundamental aspect of the system is that  students can continue
requesting and answering \textit{ad infinitum}. They receive
immediate feedback, usually including a detailed solution (see Fig. \ref{fig:quiz}). In-class
surveys indicate that students really like this.
Naturally,  students can monitor their own progress.  Several grading
schemes can be implemented, but using the last 8 answers has been the
norm until 2013. 

An Item Allocation Algorithm (IAA) is used to choose drill items (questions) for
learning,
within each lecture.
Aspects include the desire to start with easy items and increase
difficulty with increasing grade. 
Given that iteration is known to enhance learning, the IAA also
occasionally chooses an item from earlier material (lectures).
It is likely to be useful to choose again from earlier mistakes or go
to prerequisites if there is no learning, but these have not been
investigated to date.

\begin{figure}[!h]
\centering
\includegraphics[width=11cm]{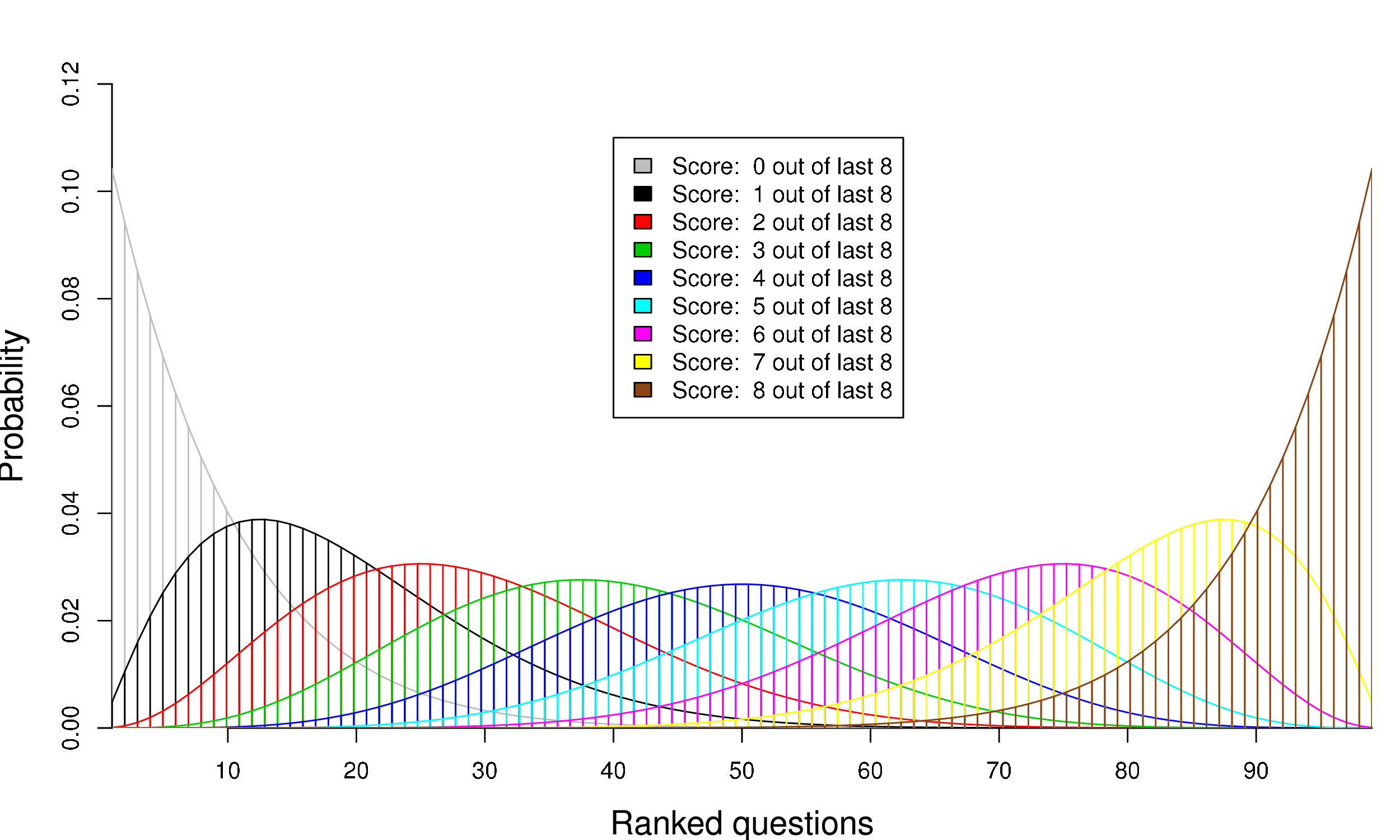}
\caption{Tutor-web probability mass function used by the item
  allocation algorithm. The x-axis indicates the ranked item
  difficulty
and the y-axis gives the probability of the next item, where the
p.m.f. choses depends on the grade of the student.}
\label{fig:iaa}
\end{figure}

The IAA is simply implemented as a probability mass function
(p.m.f., Fig. \ref{fig:iaa}), which is a function of difficulty. In addition, the
p.m.f. depends on the grade, thus implementing personalized education
appropriate for the student in question.

\begin{figure}[!h]
\begin{minipage}{0.45\textwidth}
\includegraphics[scale = 0.3,angle=-90]{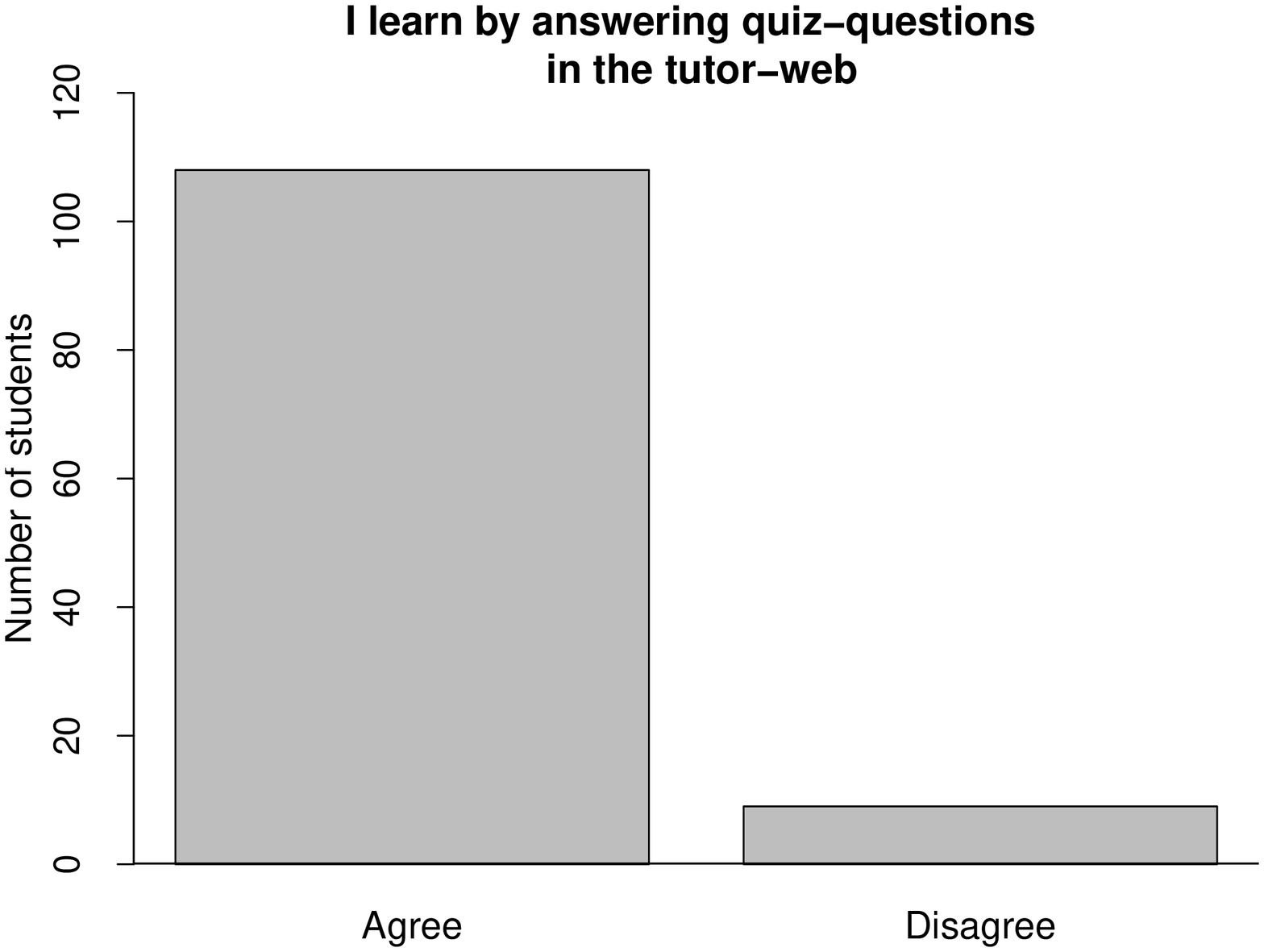}
\end{minipage}
\begin{minipage}{0.45\textwidth}
\includegraphics[scale = 0.3,angle=-90]{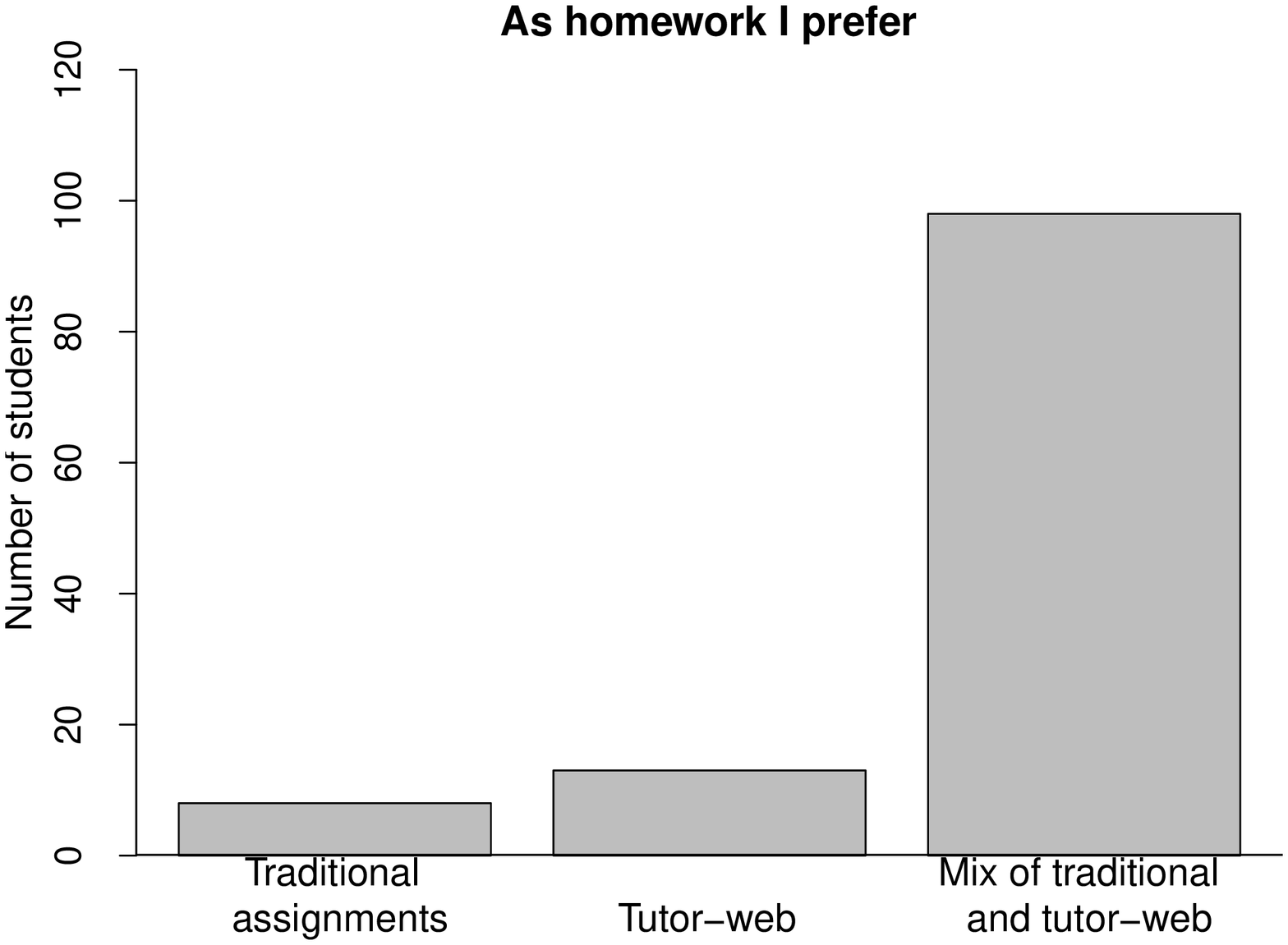}
\end{minipage}
\caption{Student satisfaction survey results. Note how the tendency to
  like web-assisted methods (left panel) does NOT imply that regular
  homework can be dropped (right panel).}
\label{fig:survey}
\end{figure}

Student surveys are conducted in most courses using the system. A
typical example of results is given in Fig. \ref{fig:survey}.Although it is useful to know that students appreciate a drilling
system, more concrete evidence is needed in order to justify its use.
One such is provided using an experimental design which compared
groups of students using the system or using traditional homework in a
crossover design \cite{jonsdottir11enhanced}. The basic conclusion from
this experiment was that the difference between the groups was
insignificant, both statistically and from the point of view that the
confidence interval for the two groups was very tight. It follows that
the system can be used to reduce regular homework considerable, but
not replace it completely (cf. Fig. \ref{fig:survey}).

\section{Monitoring students}
\begin{figure}[!h]
\includegraphics[width=11cm,angle=-90]{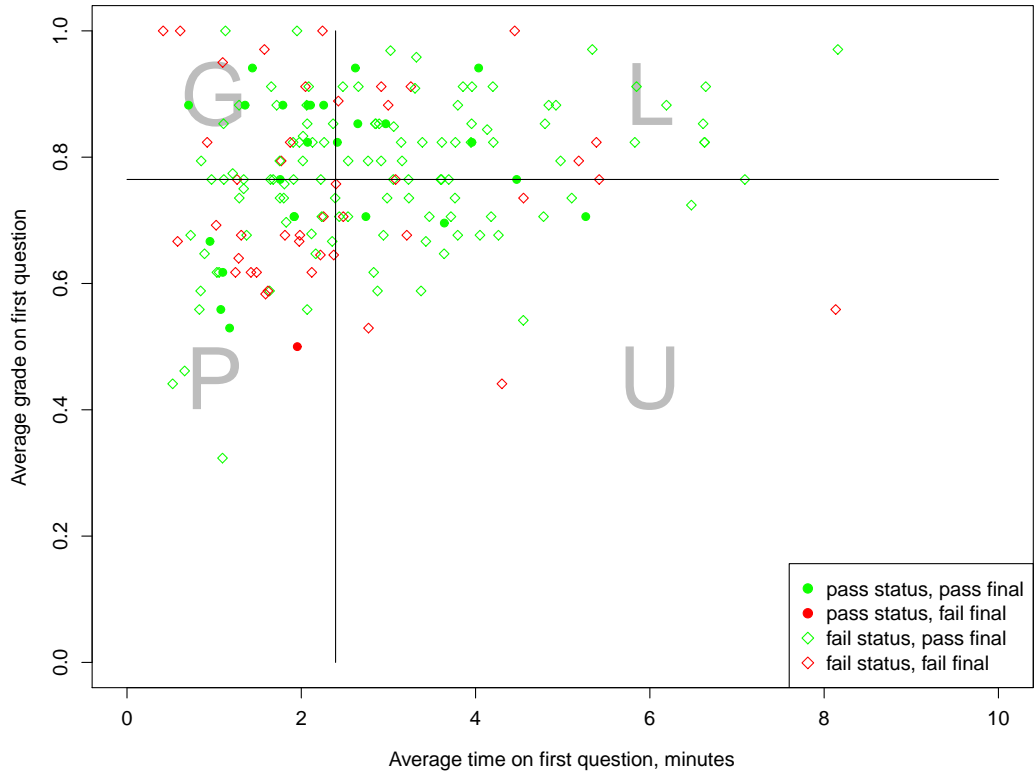}
\caption{Plot of average grade and timing for first item request within each lecture. 
Vertical and horizontal lines indicate classification of students according to time and grade (using medians). 
Color: green/red=Pass/Fail on final exam. 
Shape: Circle/Diamond=Pass/Fail on status exam. }
\label{fig:g1class}
\end{figure}

Consider next the data available to the system and how this may relate
to the actual knowledge, as determined by exams, either an initial
status exam or a final exam. A calculus course with data for 182
students is used for these analyses for the remainder of this paper. 
A status exam was submitted in the second week of the course. The problems on the exam covered numbers and functions, basic algebra, equation of a straight line, trigonometric, differentiation and integration, vectors and complex numbers. The performance on the status exam was poor with an average score of 35\%. Students were also evaluated multiple times during the semester, and
monitored continuously using the tutor-web. In the following, 
summaries of the tutor-web grade and response times along with grades from an initial status
exam and the final exam are used.
In the tutor-web, the response time for each item is measured, along with a 0/1-grade. The items are grouped in lectures as described in section \ref{introduction}, with 34 lectures belonging to this particular course. As an example, consider the average grade and average time spent on the
first item in each lecture. This provides 182 pairs. Each of these can
now be labelled in 4 ways, according to whether the student passed the
status exam and/or the final exam. These results are given in Fig.  \ref{fig:g1class}.
Notice how it is not at all clear from the figure whether
there is a link between t-w performance and grades on either exam.

A simple linear regression of grade improvement, defined as the change in grade from the status exam to the final exams, on the grade and the time used per item within the tutor-web reveals that those are important variables, but relationships to performance on exams may be nontrivial. For example, one would expect the time taken to solve a problem to be a complex combination
of the student's expertise and diligence. Thus a ``\textbf{G}ood''
student should be able to solve a problem quickly and get it right,
but the ``diligent'' hard-working \textbf{L}earner who may not know
the material very well may take longer to get the right answer. A
guessing (\textbf{P}oor) student will not take long to get the wrong
answer. The remaining (\textbf{U}nclassified) apparent non-learning
students take long to get the wrong answer. This \textbf{GLUP}
classification is derived from Fig.  \ref{fig:g1class} and used below.

\section{Relating on-line monitoring results to other performance measures}

\subsection{Relating on-line monitoring results to learning}
Although there is no trivial grouping seen in the figure, consider
using the \textbf{GLUP} - classification to predict actual learning,
or ``improvement'', using a regular ANOVA. The ``improvement'' is
defined as the change in grade from the status to the final exams where the grade of the exams has been scaled to be on the interval from 0 to 100. The ANOVA was performed using the \texttt{lm} function in R \citep{rcore}. The results are shown in Table \ref{tbl:classreg1a}.

% latex table generated in R 3.0.1 by xtable 1.7-1 package
% Thu Sep 26 14:33:09 2013
\begin{table}[ht]
\centering
\begin{tabular}{rrrrr}
  \hline
 & Estimate & Std. Error & t value & Pr($>$$|$t$|$) \\ 
  \hline
(Intercept) & 9.5582 & 2.7223 & 3.51 & 0.0006 \\ 
  class1G & 6.9671 & 4.3282 & 1.61 & 0.1092 \\ 
  class1L & 11.4772 & 3.9247 & 2.92 & 0.0039 \\ 
  class1U & 9.0109 & 4.1953 & 2.15 & 0.0331 \\ 
   \hline
\end{tabular}
\caption{Predicting improvement (final-status) from GLUP classification.} 
\label{tbl:classreg1a}
\end{table}

%\begin{verbatim}
%lm(formula = improve ~ class1 + I(Tn > T1), data = dat.tw, subset = dat.tw$last8 >     0)
%               Estimate Std. Error t value Pr(>|t|)   
%(Intercept)       6.711      2.862   2.344  0.02017 *        <- NB: P are the baseline
%class1G           7.472      4.253   1.757  0.08064 . 
%class1L          12.907      3.887   3.321  0.00109 **       <- Learners: Greatest point increase!
%class1U          11.858      4.244   2.794  0.00578 **
%I(Tn > T1)TRUE   12.047      4.339   2.776  0.00609 **       <- NB: diligence - not expertise gained
%
%Residual standard error: 19.82 on 177 degrees of freedom
%Multiple R-squared: 0.08997,	Adjusted R-squared: 0.06941 
%F-statistic: 4.375 on 4 and 177 DF,  p-value: 0.002132 
%
%\end{verbatim}

In this linear model the \textbf{P}oor students form a baseline and the estimates for the other groups can be interpreted as gain in improvement. It is therefore seen that all the other groups  perform better on average than the baseline.

The main results from this analysis are that the point estimate for
the poor performers is the lowest among the four groups. The greatest
increase from P is amongst the learners, L, but this is not
significantly different from e.g. the Good students. It is interesting
to note that the unclassified group (U) shows considerably (and
significantly) more improvement than the poor performers (P).  The
only difference in their classification is the average amount of time
spent on the items. Thus, although both groups perform poorly at the
outset, those who spent more time on each
item outperformed the other group by quite a bit on average in terms
of improvement.

\subsection{Linking to absolute performance}
Predicting the improvement during a semester, or the ``value-added''
is done directly above by fitting to the improvement in grade, from
the initial status exam to the final exam. For several reasons it is
also of interest to consider predictions of the final exam grade
(\texttt{finalG}) directly, including the status exam as a regular explanatory
variable (\texttt{statusG}). Many variables can in principle be defined and used. 
Here the average grades from different stages
within the tutor-web are included (\texttt{g1}, \texttt{g5} and \texttt{gn}, \texttt{gn} being the grade on the last item requested in the lecture), as is the average
time spent per item at different points (\texttt{T1}, \texttt{T5} and \texttt{Tn}), the squared time spent per item (\texttt{T1.2}, \texttt{T5.2} and \texttt{Tn.2}), an indicator variable of whether students spend more or less time on the last (usually most difficult) item compared with the first one (\texttt{Tn>T1}), the GLUP class variable (\texttt{class1}), number of items requested (\texttt{twnattl}) and finally the squared number of items requested (\texttt{twnattl2}). The model was fitted using the \texttt{lm} function and reduced using the \texttt{step} function in R \cite{rcore}. The result is shown in Table \ref{tbl:AICfinalsmall}.

% latex table generated in R 3.0.1 by xtable 1.7-1 package
% Thu Sep 26 15:40:11 2013
\begin{table}[ht]
\centering
\begin{tabular}{rrrrr}
  \hline
 & Estimate & Std. Error & t value & Pr($>$$|$t$|$) \\ 
  \hline
(Intercept) & -46.7506 & 11.0469 & -4.23 & 0.0000 \\ 
  twnattl & 2.4488 & 0.7809 & 3.14 & 0.0020 \\ 
  statusG & 0.5211 & 0.0609 & 8.55 & 0.0000 \\ 
  g5 & 54.3603 & 10.1337 & 5.36 & 0.0000 \\ 
  T5 & 6.0022 & 3.6711 & 1.63 & 0.1039 \\ 
  Tn & 2.9232 & 2.0771 & 1.41 & 0.1611 \\ 
  `Tn$>$T1`TRUE & 10.3281 & 4.2538 & 2.43 & 0.0162 \\ 
  twnattl2 & -0.0462 & 0.0182 & -2.54 & 0.0119 \\ 
  T5.2 & -1.0496 & 0.6013 & -1.75 & 0.0826 \\ 
   \hline
\end{tabular}
\caption{Final model selected using the AIC.} 
\label{tbl:AICfinalsmall}
\end{table}

%\begin{verbatim}
%             Estimate Std. Error t value Pr(>|t|)    
%(Intercept) -37.88670   10.41016  -3.639 0.000361 ***
%sperf         0.46389    0.05962   7.780 6.40e-13 ***  <- status exam performance
%g5           39.53381   11.01135   3.590 0.000431 ***  <- tw performance
%twnattl       2.87598    0.73784   3.898 0.000139 ***  <- negative linear in 2003; now quadratic
%twnattl2     -0.05652    0.01700  -3.326 0.001078 **   <- always significant!
%class1G      13.12982    4.36063   3.011 0.002996 **   <- "P" is reference -- "GLU" are all better
%class1L      18.46207    4.03010   4.581 8.85e-06 ***
%class1U      15.11620    3.78986   3.989 9.82e-05 ***
%sloppyTRUE   18.76316    4.91053   3.821 0.000185 ***  <- "sloppy" is subgroup of P; high on sperf
%`Tn>T1`TRUE  13.41885    3.54302   3.787 0.000210 ***  <- diligence - not expertise! -- like T1, T5, Tg0, Tn
%
%Multiple R-squared: 0.5594,	Adjusted R-squared: 0.5363 
%\end{verbatim}

Of the variables selected here, one has a slightly different status 
from the others: \texttt{statusG} is defined on data outside the tutor-web whereas other variables are
defined completely with the on-line learning system.

As can be seen in the table, the GLUP class variable is not significant when included with the grade and time at different stages, which is not surprising since the classification is defined by those variables. It should also be noted that the squared number of attempts is significant, implying that there tends to be a reduction in grade for students who give more
than 27 answers on average (per lecture). Earlier attempts at
quantification of the effect of the number of attempt have given mixed
output. For example, one might surmise that the number of attempts is
like the time spent per item, i.e. be a measure of diligence, but
there are also guessers and in fact the analyses in
\cite{stefansson2009jsm} showed a net negative linear relationship
with the number of attempts.  The greater number of students in the
present study may be the reason why it apears to be possible to
accomodate both effects using a quadratic response curve.

Note also how the effect of the time spent per item is positive (both \texttt{T5} and \texttt{Tn}), i.e. the longer the student spends on an item the higher the
final grade.  As above, this is a measure of the effect of
``diligence''.  
Finally note that the squared time spent on the 5th item was selected in this model. 
The point estimate corresponds to a reduced performance for students who use on average
more than $T_5^*=3$ minutes on the fifth item.

\section{Conclusions}
It is clear from a number of student surveys, that students from
Iceland to Kenya like an on-line drilling system, they feel they learn
from it and, based on the results given here, one can statistically
demonstrate this learning.

Research reported elsewhere \cite{jonsdottir11enhanced} imples that student learning
is almost the same, regardless of whether an on-line system or traditional
homework is used. Since the in-class
surveys consistently indicate that students prefer to also get graded homework, 
it is not possible to replace all homework by computerized drills, but one can 
easily  replace half the homework by on-line multiple-choice questions.

Since the instructor can make the drills form a part of the final grade and
can set minimum return requirements as criterion for passing, this gives considerable
potential for changes in emphases or reductions in instructor workload.

\section{Discussion: Avenues of research}

Applications of the tutor-web system have varied in student
requirements as  formal requirement are set by the instructors, not the
system.  The above results imply that it may be beneficial to
incorporate features which drive the students towards certain
behavior or performance.

In the course studied here, as well as in other courses where this
system have been tested \citep[e.g.][]{stefansson2004twe}
students tend to work towards a fairly high grade (median
$g_n=0.94$ and median of last 8 is 0.92 in the present course). Hence changes to either the allocation
algorithm or the grading scheme will likely lead to a change in
behavior where the students still work towards a goal of a high
grade, assuming it is still a feasible goal. Similarly, a timeout option is also likely to lead to
changes in student behavior. A generic positive system change has benefits over
an instructor-defined criterion since it will affect all 
students at all times, not just the course in question.

The students appear to gain (in terms of exam grade) through
requesting more items (up to 27) than normally required (8 for this
course) or normally taken (median=15, upper 75\% quartile=20). It
would therefore seem reasonable to encourage an increase in the number
of items requested by students.

The current ``last 8'' internal tutor-web grade assumes incorrect
answers until at least 8 questions have been answered in a given
lecture. Most students therefore answer at least 8 questions in each
lecture. This scheme, however, implies that if the 8th answer is
incorrect after a run of 7 answers, the grade will not increase unless
a new run of 8 correct answers is obtained. Many students stop at this
stage and this behavior is contrary to the goal of positive
reinforcement. A simple change would be to use the most recent 30
answers, 
or, more generally, to use 
for grading the most recent
$$
n_g=\max ( 8,\min ( n/2,30 ) )
$$ 
answers, possibly tapered, where $n$ is the total number of answers given. This will
penalise the guesser by introducing a longer tail and simultaneously
give reduced weight to the accidental 8th incorrect response. A
next-generation mobile-web version of the tutor-web will include
multiple grading schemes, including these. This will facilitate a
simple experiment to investigate the relationship between the grading
scheme and the number of attempts per lecture.

Although the tutor-web is a significant predictor of the final grade,
it is not a very good one. For example, of the 113 students who obtain
a grade of over 90\% on the tutor-web work, 34\% do not attain a grade
of 50\% on the final exam. The main problem with this is that the
tutor-web grade is not a reliable indicator for the students
themselves. The students with full marks, 100\% on the tutor-web, have
an 83\% chance of passing the exam however.  From this it is seen
that the tutor-web grade is ``too high'' in the sense that it
indicates more knowledge than is estimated using traditional exams.
Future work therefore needs to investigate whether changes in the
grading scheme, to the effect of lowering most grades, can provide
better indicators of exam performance.

%\newpage
Another way of ``reducing the tutor-web grade'' is to include timeout
features. Such a timeout could be a function of grade, i.e. a student
can only get into a certain grade range by answering questions
correctly within certain time limits.  

\begin{figure}[!h]
\centering
\includegraphics[width=0.5\linewidth]{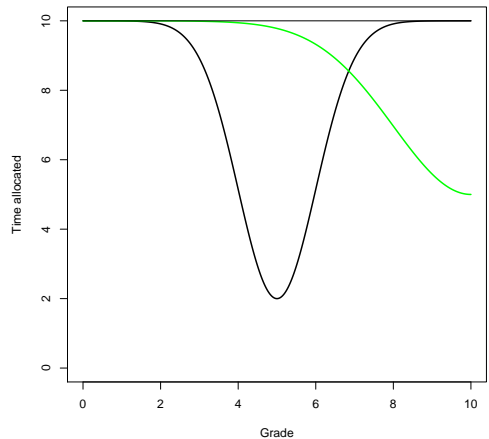}
\caption{Possible curves to define time allocated to items, as a function of grade. }
\label{fig:invdome}
\end{figure}

%\textbf{something wrong - ghostview crashes if all newpages are removed -- test by removing one at a time}
%\newpage
This will almost certainly keep students working longer within grade
intervals with a timeout and this could be used e.g. to ensure
expertise within easier items before continuing.  This approach will
also increase the number of attempts (except for the best students),
including the guessers since this will make it harder to obtain a
higher grade.  To quantify the effect of the timeout, one approach is
to focus on a single parameter in a formula such as

$$
t=a\left [1-\left ( 1 - \frac{b}{a}
             \right ) e^{-\frac{(g-g^*)^2}{2s^2}}
    \right ]
$$
%\newpage
which will give an upside-down bell-curve with an upper bound of $t=a$
and a minimum of $t=b$ at $g=g^*$. Given that the median time is about
2 minutes, one could take e.g. $a=10$, $b=2$, $g^*=5$ and $s=1$ as
initial values (cf Fig. \ref{fig:invdome}) and set up a formal experimental design by selecting
either $b$ or $g^*$ at random from within some intervals for each
student within each lecture. Performance can be evaluated
statistically either by how the number of attempts within a lecture
changes as a function of $b$ or by how the performance on an algebra
item in an exam varies as a function of $b$. This particular choice of
parameter values enforces a bottleneck where the students have to
obtain a certain level of expertise before getting above a certain
grade, upon which the timeout parameter is no longer limiting. A
different approach (using a higher $g^*$ and $s$) would be to set a
similar limit access to the higher grades. Given the complex
relationship described in this paper, between time spent on each item
and subsequent performance, it is not trivial to predict the full
effect of any timeout parameter settings.

%\newpage
Finally, since the Poor students (in the GLUP classification) are the
poorest performers by all measures, one needs to consider methods to
move these students into the otherwise Unclassified group, who spend
more time on each item. When students have answered a question the
system provides a detailed explanation of how the answer is obtained
(most items have such explanations).  A possible method to slow
these students down is therefore to use pop-ups, such as a warning
when a student has answered incorrectly and clearly asks for the next
item without first reading the explanation. The net effect of this can
easily be tested by randomly assigning such stop-signs to half the
P-students and evaluating whether there is a statistical difference in
how they move out of the P group.

\section{Acknowledgements}
The tutor-web project has been supported by the Marine Research
Institute of Iceland, the United Nations University and the University
of Iceland. Numerous authors have contributed educational material to
the system. The system is based on Plone
\footnote{http://www.plone.org} and educational material is mainly
written in \LaTeX.  Examples and plots are mostly driven by R
\citep{rcore}.  The current Plone version of the tutor-web has been
implemented by Audbjorg Jakobsdottir but numerous computer programmers
have contributed pieces of code during the lifetime of the project.

The calculus material content used as a basis for this research has
been developed by numerous faculty and students at the University of
Iceland.

\newpage
%\section{References*}
%\bibliography{../../sources/general} % first this

              % then this

\end{document}